\documentclass[aip,amsmath,amssymb,reprint]{revtex4-1}
\usepackage{hyperref}
\usepackage{graphicx} 
\usepackage[dvips]{epsfig}
\usepackage{xcolor}
\usepackage[normalem]{ulem}
\newcommand{\beq}{\begin{equation}}
\newcommand{\eeq}{\end{equation}} 
\newcommand{\beqa}{\begin{eqnarray}}
\newcommand{\eeqa}{\end{eqnarray}}
\newcommand{\ba}{\begin{array}}
\newcommand{\ea}{\end{array}}

\begin{document}

\title{Concurrent processes in the time-resolved solvation and Coulomb
  ejection of sodium ions in helium nanodroplets}

\author{Florent Calvo}
\email{florent.calvo@univ-grenoble-alpes.fr}
\affiliation{Universit\'e Grenoble Alpes, CNRS, LIPHY, F38000
  Grenoble, France}

\begin{abstract} 
Recent pump-probe experiments [Albrechtsen {\em et al.}, Nature {\bf
    623}, 319 (2023)] have explored the gradual solvation of sodium
cations in contact with helium nanodroplets, using a fully solvated
xenon atom as a probe exerting a repulsive interaction after its own
ionization. In this Communication we computationally examine by means
of atomistic ring-polymer molecular dynamics the mechanisms of
successive ionizations, shell formation, and Coulomb ejection that all
take place within tens of picoseconds, and show that their interplay
subtly depends on the time delay between the two ionizations but also
on the droplet size. The possibility of forming solvated Na$^+$Xe
non-covalent complexes under a few tens of picoseconds in such
experiments is ruled out based on fragment distributions.
\end{abstract} 
\date{\today}

\maketitle

Ion solvation is an essential process in chemical physics, involved in
very diverse areas such as atmospheric and biological
sciences.\cite{marcus} While its main features at equilibrium are well
captured in bulk environments through macroscopic concepts, monitoring
the stepwise formation of solvation shells has remained a formidable
challenge for experimentalists until the last decades. In a recent
laser experiment, helium nanodroplets (HNDs) were used as a solvation
medium, hosting impurities whose different behavior could be exploited
to monitor, in real time, gradual shell formation around the first
ionized impurity.\cite{albrechtsen} HNDs have been employed since the
1990s as a cryogenic medium enabling high resolution spectroscopic
studies of various atomic or molecular
dopants,\cite{goyal,toennies04,choi06,tigstie07} but also, for
themselves, as a probe of strongly collective quantum effects such as
superfluidity. In the aforementioned experiment, the helium droplets
were initially doped with both heliophobic (sodium) and heliophilic
(xenon) atoms, subsequently ionized with ultrashort lasers and the
ejection of the partially solvated alkali ion caused by the Coulomb
repulsive force between the ions was monitored by mass spectrometry
coupled with velocity map imaging. A wealth of invaluable details were
thus inferred from those measurements about the rate at which helium
atoms are initially captured by the newly formed ion.

The experiment of Stapelfeldt and coworkers, which provides a rather
direct observation of time-resolved ion solvation in helium droplets,
follows earlier attempts at monitoring the same process from
spectroscopy.\cite{leal14} It was complemented by dedicated
computational modeling using a continuum description of the droplet
and its dynamical evolution upon sodium ionization based on the
time-dependent helium density-functional theory (TD-DFT)
approach\cite{dalfovo95} and semiclassical dynamics. This model
confirmed the generic picture of fast capture of the helium atom by
the newborn ion under the observed time scale but with limited
submersion for 2000-atom droplets. Applied to ions other than Na$^+$,
the TD-DFT method found the extent of submersion to be quite dependent
on the ion itself,\cite{leal14,mateo14,garcia24} some ions like Cs$^+$
being notably predicted to not even sink into 1000-atom
droplets,\cite{leal14} although they readily do into
He$_{2000}$.\cite{garcia24} In superfluid droplets, successful
submersion is also expected to produce an oscillatory motion with a
period of several tens of picoseconds due to the reduced energy
dissipation once the snowball is formed.\cite{mateo14}

In an earlier contribution,\cite{submersion} a fully atomistic
approach was employed to model the submersion process of Na$^+$ and
Na$_2^+$ into small droplets containing at most several hundreds of
helium atoms, the solvation and sinking time scales being inferred
from the variations of appropriate time autocorrelation functions. The
ring-polymer molecular dynamics (RPMD) method, which is based on
Feynman's description of quantum statistics extended to the evaluation
of time correlation functions,\cite{craig04} was employed out of
equilibrium\cite{richardson09,menzeleev10,hele13,welsch16} to capture
the same mechanisms at play during the early solvation process. It was
there suggested that the solvation shell is formed concomitantly with
the sinking of the sodium cation to the center of the droplet,
predicted to take about 15~ps for the 500-atom droplet, both the
solvation and submersion times extending markedly for the Na$_2^+$
impurity.\cite{submersion} No oscillatory motion was found under the
normal fluid conditions assumed in this work.

In the present Communication, and in closer relation to the experiment
of Albrechtsen {\em et al.},\cite{albrechtsen} we extend this
computational effort to more realistic droplet sizes of 1000--5000
atoms and address the specific role of the xenon dopant, before and
after its ionization. The simulations consist of three stages. First,
the droplets containing neutral sodium and xenon impurities with
prescribed numbers of helium atoms are thermalized using standard
path-integral MD with massive Nos\'e-Hoover thermostatting. At time
zero, the sodium atom is suddenly ionized, no complex multiphoton
processes being at play for sodium in the
experiment.\cite{albrechtsen} The dynamics following sudden ionization
is monitored in real time by RPMD, including the local environment of
the ion but also its location relative to the still neutral xenon
atom. The third stage consists in ionizing the xenon dopant, a process
which in the experiment is multiphotonic owing to the much higher
ionization potential of this element. Because we mainly monitor the
sodium cation and are not especially interested in the local
rearrangement in the vicinity of the xenon atom undergoing ionization,
we simplify the modeling and assume again that Xe$^+$ is formed
suddenly at a controlled time delay $\delta t$ after the sodium atom.

The PIMD and RPMD methodologies employed here are well known to be
approximate in several respects, starting with the neglect of exchange
effects that are important in helium droplets below the superfluid
temperature of 0.4~K. Extensions to incorporate such effects have been
proposed in the literature \cite{hirshberg19,brieuc20} but they remain
involved or simply incompatible with any dynamical treatment of the
nuclear degrees of freedom, a feature which is precisely essential
here. Acknowledging this limitation, and also alleviating a
significant portion of the computational cost, the initial temperature
of the droplets was fixed above 0.4~K, namely 1~K for He$_{1000}$ and
He$_{2000}$ droplets and 2~K for He$_{5000}$, leading to normal fluids
in which nuclear delocalization is accounted for. The second major
approximation is the use of RPMD for out-of-equilibrium dynamics,
since the system changes potential energy surfaces twice with each
ionization process. Despite important progress in path-integral
methods for time correlation functions, also away from equilibrium,
\cite{richardson09,menzeleev10,hele13,welsch16,calvo14} it is clear
that the quantum dynamics they describe remains valid at short times
and that only semi-quantitative details can be reasonably described
with this approach.

To cope with the rather large sizes of the droplets, simplified but
efficient potentials were used to describe the interactions in the
NaXe@He$_n$, Na$^+$Xe@He$_n$, and Na$^+$Xe$^+$@He$_n$ systems, with
parameters chosen to reproduce available reference data. More
precisely, simple Lennard-Jones (LJ) interactions were employed in the
neutral system for all pairs, namely He-He, Na-He, Xe-He, and Na-Xe.
In the singly cationic case (Na$^+$ and neutral Xe), the interactions
with the ions comprised an additional contribution accounting for the
ion-induced dipole, either as a simple pair term $(-\alpha_{\rm
  Xe}/2r^2)$ for the interaction between Na$^+$ and He, or as a more
accurate vectorial polarization term $-\alpha_{\rm He}\vec E_i^2/2$ on
each helium atom $i$, $\vec E_i$ being the local electric field
created by the sodium ion. In both contributions, $\alpha_{\rm
  Xe}=4.010$~\AA$^3$ and $\alpha_{\rm He}=0.205$~\AA$^3$ denote the
isotropic atomic polarizabilities of the neutral xenon and helium
atoms, respectively. After ionization of the xenon atom, we set a
simple Coulombic repulsion between the two ions, and the electric
field $\vec E_i$ on each helium atom now incorporates the contribution
of the xenon ion as well. Keeping the isotropic polarizabilities as
fixed parameters, all LJ parameters were adjusted to reproduce the
relevant pair interactions. They are tabulated as Supplementary
Electronic Material, together with references to the original data
used to fit these interactions, and are illustrated in Fig.~S1. Other
computational details of the present simulations are also provided as
Supplementary material.

Unsurprisingly, and owing to the highly polarizable character of the
heavy xenon element, the Na$^+$Xe non-covalent complex is the most
strongly bound of all pairs involved in this experimental situation,
reaching almost 0.3~eV, to be compared with the few meVs of all
neutral pairs, and a few tens of meV for other ionic pairs Na$^+$-He
and Xe$^+$-He. In neutral state, Na-Xe is also quite more strongly
bound than any of the pairs involving helium, but this attraction is
not sufficient to bring the two neutral impurities in contact with
each other during the initial equilibration period, a result that is
consistent with earlier TD-DFT simulations of comparable
systems.\cite{poms}

The formation of a helium shell around the sodium ion was monitored in
the RPMD simulations by a simple measure of its coordination number
$\langle N_C\rangle$ averaged over polymer beads through appropriate
cut-off functions,
\begin{equation}
  N_c[{\bf R}(t)] = \sum_{i\in\rm HND} f_c(r_{\rm Na-He}^{(i)}),
  \label{eq:nc}
\end{equation}
where ${\bf R}(t)$ denotes the set of Cartesian coordinates of the
Na$^+$Xe@He$_N$ system, $r_{\rm Na-He}^{(i)}$ the distance between the
sodium ion and the $i^{\rm th}$ helium atom, and $f_c(t)$ a function
such that $f_c(r<r_1)=1$, $f_c(r>r_2)=0$, being interpolated smoothly
between those two values using a fifth-order polynomial. Here
$r_1=3.84$~\AA\ and $r_2=4.11$~\AA\ were chosen to ensure that a
helium atom does not belong to the shell if sodium is still in neutral
form. $\langle N_c\rangle$ is sensitive to the first helium shell only
and does not measure the total number of helium atoms captured by the
ejected ion. Here, the limited statistics of 10 independent
trajectories prevents from attempting an accurate determination of the
solvation time from the time autocorrelation function associated with
$\langle N_c\rangle$, hence we resort to the simple time variations of
this direct quantity.

\begin{figure}[htb]
  \centerline{\includegraphics[width=9cm,clip]{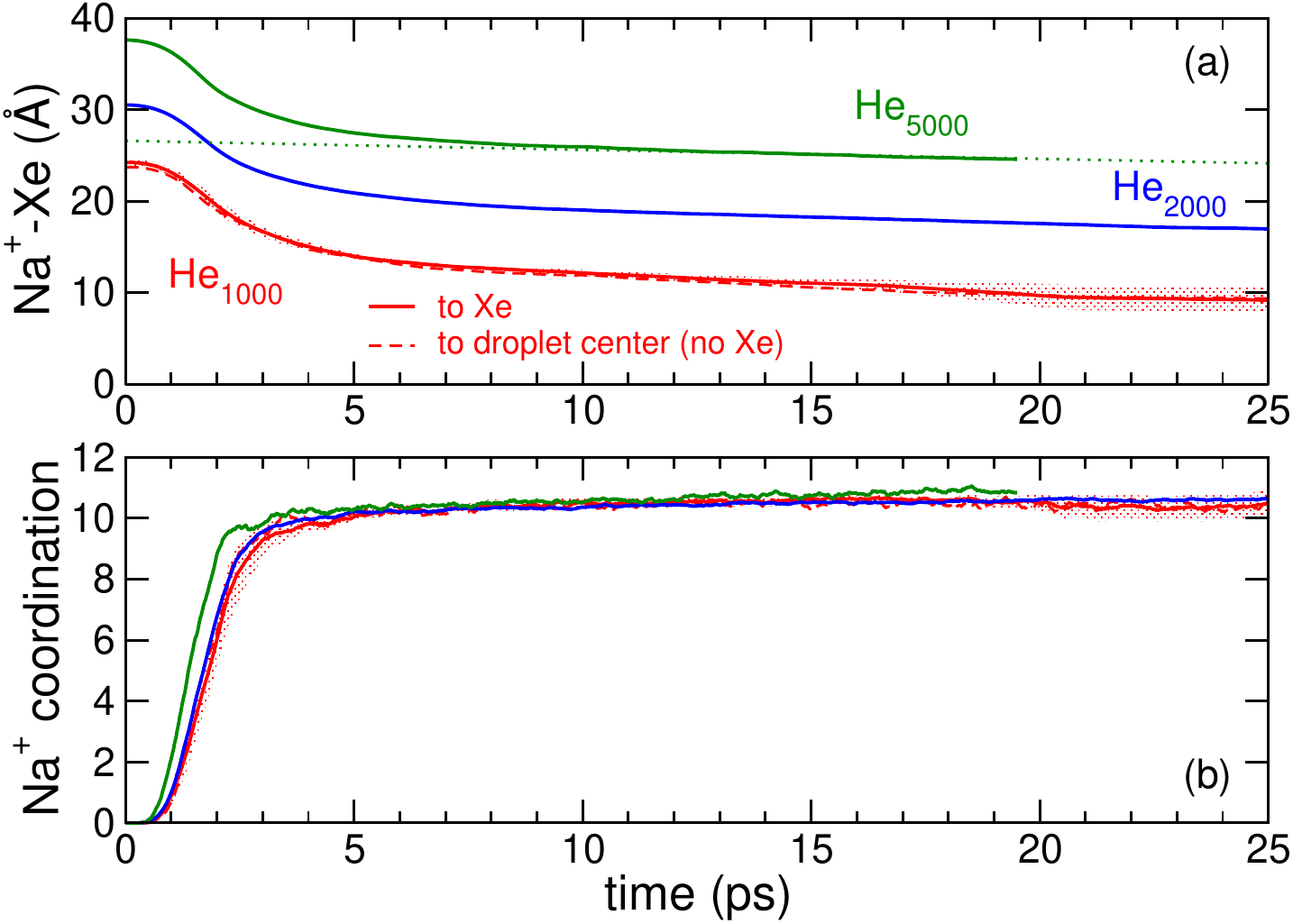}}
  \caption{(a) Average Na$^+$-Xe distance, as a function of time after
    ionization of the sodium impurity, for droplets containing 1000,
    2000, and 5000 helium atoms, averaged over 10 independent RPMD
    trajectories. For He$_{1000}$, a range of uncertainty is shown
    from the fluctuations among the 10 independent trajectories. The
    average distance between Na$^+$ and the droplet center is also
    shown for He$_{1000}$ droplets lacking the xenon dopant. The
    dotted line for He$_{5000}$ suggests a linear extrapolation at
    longer times; (b) Corresponding variations of the average
    coordination number of the Na$^+$ ion, with fluctuations only
    shown for the 1000-atom droplets.}
  \label{fig:naionization}
\end{figure}

Fig.~\ref{fig:naionization} shows the time variations of the distance
between the two impurities, after sodium is ionized at time $t=0$, and
the corresponding coordination number $\langle N_c\rangle$, for the
three droplet sizes of 1000, 2000, and 5000 atoms.
For He$_{1000}$ droplets, additional simulations were performed
without xenon dopant, and Fig.~\ref{fig:naionization} also shows the
variations of the average distance between Na$^+$ and the center of
mass of the droplet upon ionization. The results are barely
distinguishable from those obtained for the Xe@He$_{1000}$ droplet,
indicating that the xenon dopant exerts no significant attraction on
the sodium cation at this early stage.

The decrease of the Na$^+$-Xe distance is monotonic but proceeds
through two main stages. During the first picosecond, the system
adjusts to its new potential energy surface and the distance decreases
slowly but quadratically with time, as expected from semiclassical
dynamics where the coordinate of interest has no initial velocity.
During the subsequent 3--4 ps, the Na$^+$-Xe distance decreases faster
and the extent of sinking reaches about 10~\AA, independently of
droplet size. Submersion occurs concomitantly with the formation of
the solvation shell of 10 helium atoms around the sodium cation, but
it is interesting to note that the shell is formed prior to the
10~\AA\ submersion, indicating that initially it is the helium atoms
that lied in the vicinity of the neutral sodium atom that actually got
pulled by the newly formed cation through ion-induced dipole forces,
before the ion itself started to sink. This mechanism is also clearly
present in the TD-DFT simulations of Albrechtsen {\em et al.}
\cite{albrechtsen}.

Droplets of a few hundreds of atoms have a radius comparable or
smaller to the extent of submersion found here of about 10~\AA\ after
5~ps, and this explains why our earlier RPMD simulations
\cite{submersion} found that submersion was complete with the ion
reaching the droplet center within about 10 ps. Here, submersion
remains incomplete even for He$_{1000}$ droplets, except for one
trajectory in the sample, where the sodium cation meets the xenon
dopant about 15-20~ps after ionization, as seen on the increasing
fluctuations found for the Na$^+$-Xe distance in
Fig.~\ref{fig:naionization}(a). For this trajectory, the formation of
the Na$^+$-Xe non-covalent bond leads to a partial decrease in the
size of the solvation shell, seen together with increasing statistical
fluctuations in Fig.~\ref{fig:naionization}(b) at times $t>15$~ps.

The very slow but steady decrease in the Na$^+$-Xe distance can be
roughly extrapolated linearly to evaluate the time it would take for
complete submersion, that is formation of the Na$^+$-Xe non-covalent
bond, for all droplet sizes simulated here. Employing the same linear
slope extracted for the 5000-atom droplet highlighted with a dotted
line in Fig.~\ref{fig:naionization}(a), we find times of 70, 100, and
165 picoseconds for He$_{1000}$, He$_{2000}$, and He$_{5000}$,
respectively. More generally, for an arbitrary He$_N$ droplet, the
same procedure can be used to predict that the submersion time should
vary approximately as $t_s \simeq t_0+\alpha N^{1/3}$, with
fitted parameters $t_0=-71.25$~ps and $\alpha=13.89$~ps.

\begin{figure}[htb]
  \centerline{\includegraphics[width=9cm,clip]{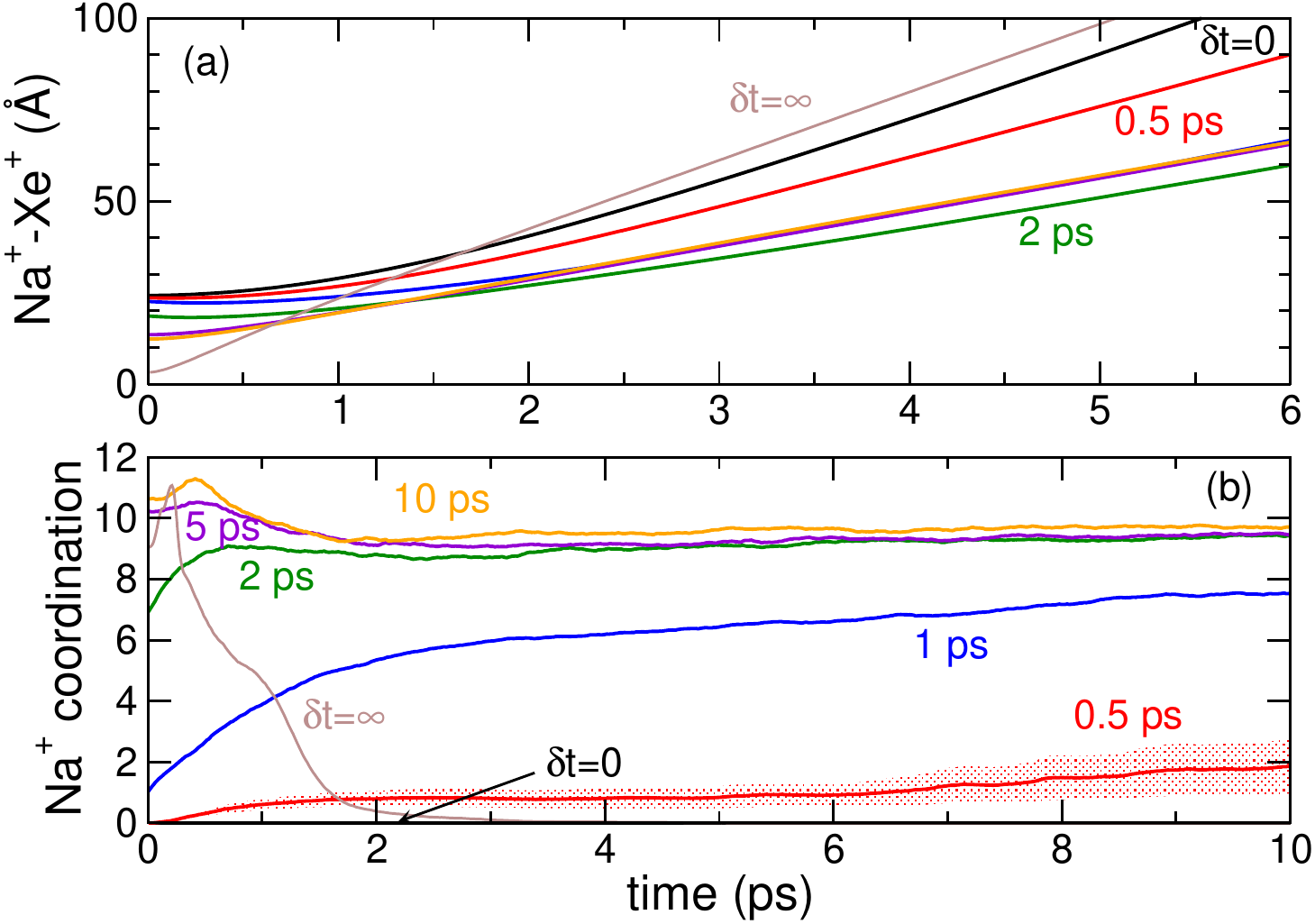}}
  \caption{(a) Average distance between the sodium and xenon ions in
    He$_{1000}$ droplets, as a function of time after the xenon dopant
    was ionized with a delay of $\delta t$ after the sodium atom; (b)
    Corresponding variations of the coordination number of the Na$^+$
    ion. A delay time of $\infty$ means that the system was prepared
    as the fully solvated Na$^+$Xe complex at equilibrium.}
  \label{fig:xeionization_1000}
\end{figure}

The dynamics following the second ionization of the xenon dopant is
next examined. Fig.~\ref{fig:xeionization_1000} shows essentially the
same properties as Fig.~\ref{fig:naionization}, but for the 1000-atom
droplets only and for different values of the time delay $\delta t$
between ionization of the two impurities, $\delta t=0$, 0.5, 1, 2, 5,
and 10~ps.
Under the conditions of simultaneous ionizations ($\delta t=0$), no
helium atom is captured by the ejected sodium ion, only a minor
fraction ($\langle N_c\rangle \sim 0.1$) being noticed about 1--2~ps
after this double ionization, indicating that for one trajectory only,
one helium atom had the time to get pulled close enough to the sodium
cation before this ion was finally ejected, naked. If the xenon atom
is ionized 0.5~ps after the sodium atom, a few helium atoms are
definitely captured, but the statistical fluctuations in the
coordination number of Fig.~\ref{fig:xeionization_1000}(b) are quite
significant, indicating a peculiar mechanism involving delayed
capture. Some example trajectories, illustrated in Fig.~S2 (Multimedia
available online) of the electronic Supplementary Material, show that
for this specific size and at this specific delay time the Coulomb
repulsive force balances the attractive solvation (polarization) force
originating from the neighboring helium atoms, in such a way that the
cation remains in the vicinity of the droplet, without sinking
significantly [by less than $0.1$~\AA, see
  Fig.~\ref{fig:naionization}(a)] but staying long enough for the
nearest helium atoms to be pulled towards the cation, still as this
ion has turned back and begun its ejection process.

This explains the complex, two-stage behavior in the time variations
found for the coordination number at this droplet size. As the xenon
atom is ionized at longer time delays, the sodium cation is able to
coat itself with more solvent atoms before being ejected, the full
coordination shell of about 10 atoms being formed already after 2~ps.

It is interesting to compare the above situation of finite time delays
to the asymptotic case of an initially fully submersed cation. For
this problem, additional equilibrium PIMD simulations were performed
for the Na$^+$Xe non-covalent complex embedded inside the helium
droplet. The corresponding post-ionization dynamics is referred to as
$\delta t=\infty$ in Fig.~\ref{fig:xeionization_1000}. Quite
strikingly, under such assumption the sodium cation is ejected from
the droplet in a couple of picoseconds so strongly that any attached
solvent atom is vaporized.

\begin{figure}[htb]
  \centerline{\includegraphics[width=9cm,clip]{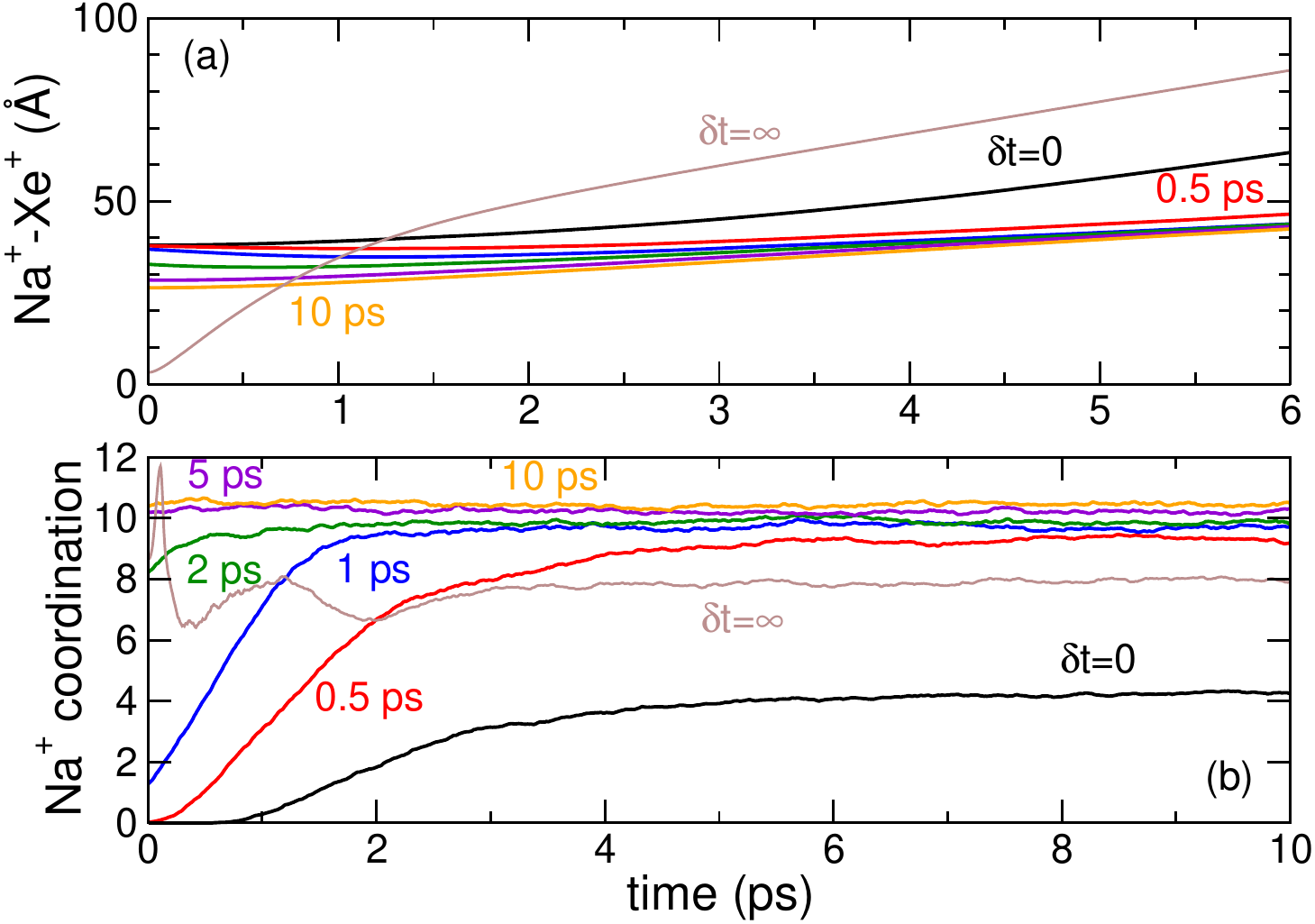}}
  \caption{(a) Average distance between the sodium and xenon ions in
    He$_{5000}$ droplets, as a function of time after the xenon dopant
    was ionized with a delay of $\delta t$ after the sodium atom; (b)
    Corresponding variations of the coordination number of the Na$^+$
    ion. A delay time of $\infty$ means that the system was prepared
    as the fully solvated Na$^+$Xe complex at equilibrium.}
  \label{fig:xeionization_5000}
\end{figure}

In the larger He$_{5000}$ droplets the mechanisms of Coulomb ejection
are much softer, as seen in Fig.~\ref{fig:xeionization_5000} where the
same properties are represented for the same time delays.
In particular, even for $\delta t=0$ the attenuation of the Coulomb
repulsion relative to the He$_{1000}$ droplet, caused by the larger
distance between the two impurities (38~\AA\ versus 24~\AA), leaves
far more time for the sodium cation to capture helium atoms,
non-monotonic effects being no longer found for $\delta t=0.5$~ps (see
Fig.~S2 in Supplementary Material). Even under the full submersion
asymptotic conditions of $\delta t=\infty$, the sodium cation still
manages to travel accross the droplet and be ejected with a partially
filled solvation shell of 8 atoms in average.

Comparing now with the experimental measurements of
Ref.~\citenum{albrechtsen}, the number of helium atoms carried with
the ejected sodium cation appears to agree reasonably well for the
He$_{5000}$ droplet, but is definitely too small for He$_{1000}$,
which we interpret as reflecting the excessively strong Coulomb
repulsion in this smaller droplet: due to the smaller droplet size,
the impulse felt by the sodium cation upon xenon ionization is about
2.5 times stronger than in He$_{5000}$.

\begin{figure}[htb]
  \centerline{\includegraphics[width=9cm,clip]{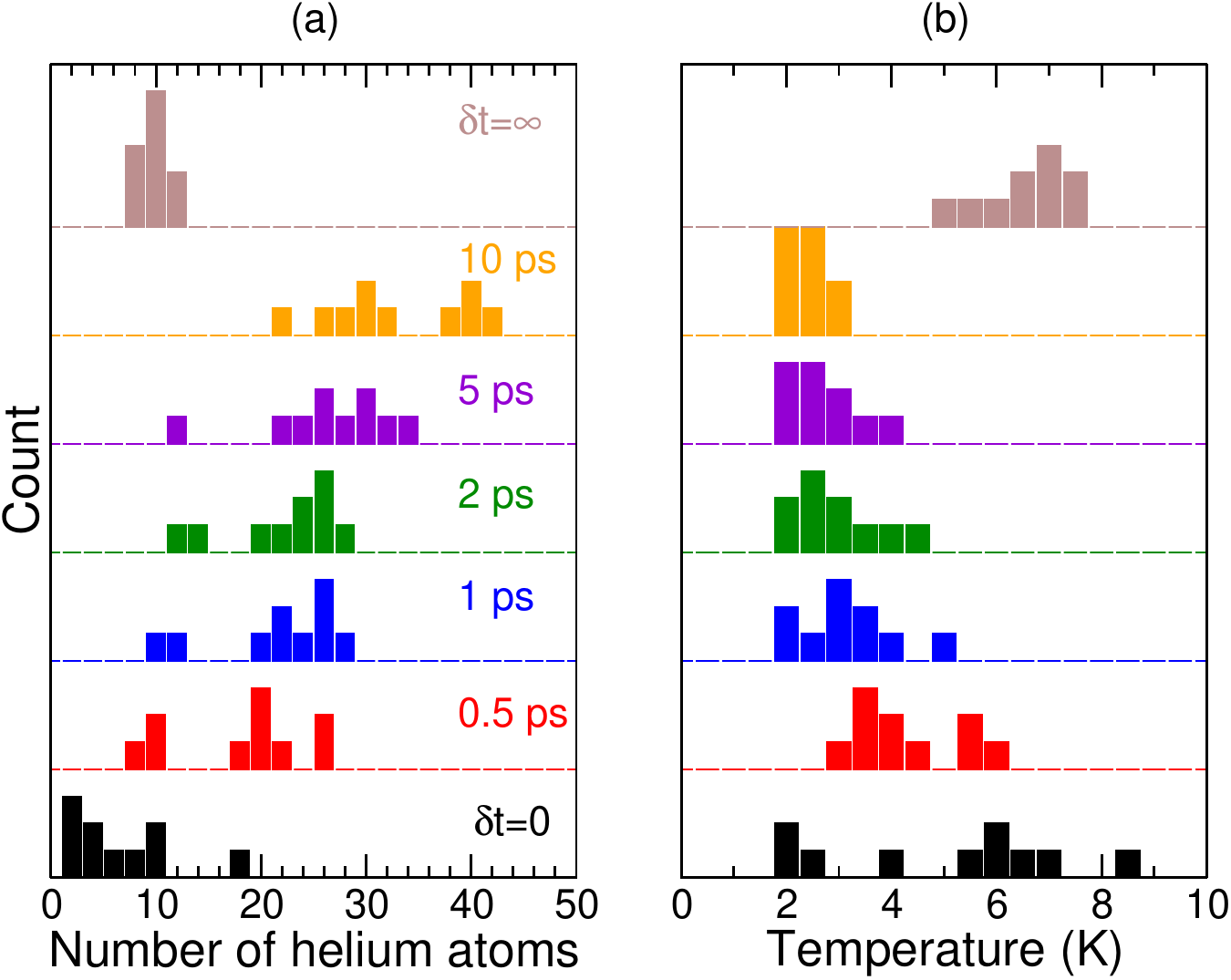}}
  \caption{(a) Distributions of the number of helium atoms carried by
    the ejected sodium ion in contact with He$_{5000}$ droplets, 20 ps
    after the xenon dopant was ionized, for different delay times
    $\delta t$ between the successive ionizations of both impurities;
    (b) Distributions of the corresponding temperatures of these
    fragments.}
  \label{fig:fragstat}
\end{figure}

Even with limited statistics, our results can be compared to the
experimental measurements, and Fig.~\ref{fig:fragstat}(a) shows the
distributions of the numbers of helium atoms carried by the sodium
cation 20~ps after ionization of the xenon atom in He$_{5000}$
droplets and for increasing time delays $\delta t$. This number was
obtained by partitioning the atoms into various fragments, defined
according to a simple connectivity criterion with a cut-off distance
of 7~\AA, distances being measured between centroids. While the
coordination number only accounts for the first solvation shell around
the cation, the total number of helium atoms attached to the cation
can thus be much larger.
For finite time delays, the fragment size distributions are relatively
broad and vary towards increasing values with increasing $\delta
t$. In contrast, the helium atoms carried away by the ejected sodium
cation initially in contact with a xenon atom are fewer and less
broadly distributed. The distributions found in the present
simulations somewhat overestimate those measured in the
experiment.\cite{albrechtsen} To solve this discrepancy, we have
extended the RPMD trajectories to the ejected fragments Na$^+$He$_p$
alone and let them evolve for 10 more picoseconds, initially removing
the overall linear momentum. A vibrational temperature could thus be
estimated from the excess kinetic energy, and the corresponding
statistical distributions are shown in Fig.~\ref{fig:fragstat}(b).

The temperatures estimated for the fragments are notably high and
increase with decreasing time delays, but are also significant in the
asymptotic limit $\delta t=\infty$. In both cases, high temperatures
indicate a solvation shell that was still not equilibrated 20~ps after
xenon ionization, either because it results from sticking collisions
to the departing cation (short $\delta t$), or because it originates
from the violent ejection of the cation through the droplet, dragging
with it only a limited number of vibrationally excited solvent
atoms. For fragments larger than a dozen atoms, the outer helium atoms
that are not tightly bound to the sodium cation are particularly prone
to statistical evaporation and we anticipate that smaller fragments
would be observed after the experimental time scale of several
hundreds of ps, in closer agreement with the measurements but also the
classical MD simulations reported in
Ref.~\citenum{albrechtsen}. Besides the ejection of the sodium cation
and its flock of helium atoms, the droplet relaxes by a number of
other mechanisms that operate at short (rearrangement of the solvation
shell), intermediate (evaporation of individual helium atoms from the
droplet), or longer times (recoil motion of the xenon cation inside
the droplet). In the Supplementary Material (Fig. S3), the two first
aspects are scrutinized in more details.

In conclusion, the picture emerging from our path-integral simulations
appears to be rather faithful to the experiment carried by Stapelfeldt
and coworkers.\cite{albrechtsen} By reproducing its main features, our
modeling shows that the fragments size distributions are not only a
signature of cation solvation itself, but also of the size of the host
droplet. The superfluid character of the droplet, which may break down
in the vicinity of the newly formed ions,\cite{gonzalez} is not a
required property although it is naturally expected to play a role in
the ionic dynamics and the propensity of the ejected cation to keep or
release helium atoms. It should also play a role in the possible
oscillatory motion that could occur at longer times,\cite{mateo14}
although the formation of the rather strongly bound Na$^+$Xe complex,
predicted here to occur after hundreds of picoseconds, should be a
further major source of energy dissipation in the system.

The nature of the alkali element should also influence the fragment
distribution, and possibly the rate at which solvent atoms are
captured,\cite{garcia24} at least through the different equilibrium
distances its neutral atom lies from the droplet before ionization, as
well as different solvation shells it can accomodate. This will be the
object of future efforts.

\section*{Supplementary Material}

Details of the interaction potentials and of the PIMD and RPMD
simulations; enhanced discussion about two-stage helium capture and
energy dissipation; selected animations.

\end{document}